\documentclass[12pt]{article}

\usepackage{pifont}
\usepackage{bm}
\usepackage{xcolor}
\usepackage{comment}
\usepackage{hyperref}
\usepackage{caption}
\usepackage{amsmath}
\usepackage{bbold}
\usepackage{etoolbox}
\usepackage{adjustbox}
\usepackage{makecell}
\usepackage{subcaption}
\usepackage{textcomp}
\usepackage[symbol]{footmisc}
\usepackage{booktabs}
\usepackage{multirow}
\usepackage{algorithm}
\usepackage{algorithmic}
 \usepackage{setspace}
 \usepackage{textcomp}
\usepackage[letterpaper,margin=1in]{geometry}
\usepackage{mathptmx}
\usepackage[T1]{fontenc}
\usepackage[utf8]{inputenc}
\usepackage{amsmath,amssymb}
\usepackage{booktabs}
\usepackage{longtable}
\usepackage{enumitem}
\usepackage{url}
\usepackage{calligra}
\usepackage{graphicx}
\usepackage{subcaption}
\usepackage[most]{tcolorbox}
\usepackage{caption}
\usepackage[numbers,sort&compress]{natbib}
\usepackage{tabularx}
\usepackage{pifont}

\newcommand{\pass}{\textcolor{green!50!black}{\ding{51}}}
\newcommand{\fail}{\textcolor{red!65!black}{\ding{55}}}

\setlist[enumerate]{leftmargin=*, itemsep=0.25em, topsep=0.4em}
\setlist[itemize]{leftmargin=*, itemsep=0.25em, topsep=0.4em}
\newcommand{\correspondingauthor}[1]{\noindent\textbf{Corresponding author:} #1\par\bigskip}

\begin{document}


\begin{center}
{\Large\bfseries Using Local Large Language Models to Build Reproducible and Differentiable ODE Parameter Estimation Pipelines}\par\vspace{1em}
{\large Saakaar Bhatnagar$^{1}$}\par\vspace{0.5em}
$^{1}$\textit{Independent Researcher}\par
\end{center}

\correspondingauthor{Saakaar Bhatnagar, \texttt{saakaar28@gmail.com}}


\section*{Abstract}
Parameter estimation for ordinary differential equation (ODE) models requires a sequence of interconnected tasks, including implementing model equations, mapping simulated quantities to experimental observations, constructing objective functions, configuring numerical solvers, and assessing fitted results. This chapter presents an agentic workflow that uses locally deployed large language models (LLMs) to assist with these tasks while retaining human oversight of scientific decisions. The workflow divides the estimation process into focused stages, each supplied with task-specific instructions and subject to validation before proceeding. Starting from a mathematical problem description and experimental data, it generates a structured configuration, a readable model implementation, and differentiable code for numerical optimization. Population-based global search is followed by gradient-based refinement using automatic differentiation through the ODE solution. Post-fit diagnostics examine residuals, optimization behavior, and local parameter sensitivity, while preserved intermediate artifacts allow users to inspect assumptions and revisit individual stages. A worked example demonstrates the complete procedure, and additional examples illustrate its application to diverse dynamical systems. The chapter provides a practical approach to constructing and revising ODE parameter-estimation pipelines in locally managed computing environments.
\section*{Key words}
 Parameter Estimation; Ordinary Differential Equations; Global Optimization; Differentiable Programming; Gradient-Based Optimization

\section{Introduction}

\subsection{Motivation}

Ordinary differential equation (ODE) models are widely used to represent biological, chemical, and physical systems, both independently and as components of larger multiphysics models. The predictive accuracy of these models depends on parameters that must often be inferred from experimental measurements. Parameter estimation is commonly formulated as an optimization problem that minimizes a discrepancy, such as a weighted sum of squared residuals, between model-predicted observables and experimental data \cite{rodriguez2006hybrid}. These observables may be state variables, their time derivatives, or other quantities derived from the ODE solution. Population-based global search methods provide derivative-free approaches to exploring the parameter space efficiently since independent candidate evaluations can be parallelized. Infeasible parameter sets or failed simulations can be handled through rejection or penalty mechanisms, provided these are incorporated into the optimization procedure. However, such methods can require many forward simulations, motivating hybrid strategies that combine global exploration with gradient-based local refinement \cite{rodriguez2006hybrid}.

These methods return a good initial guess to the problem, which can be fine-tuned using gradient based methods to obtain a more accurate fit. An additional challenge for gradient based methods is that for numerical derivatives of the solution w.r.t parameters may be highly unstable or inaccurate. This depends on the numerical properties of the ODE and the dimensionality of the estimation problem. Differentiating the discretized ODE solver using automatic differentiation, with suitable checkpointing, provides accurate parameter gradients for optimization while avoiding instabilities associated with backward reconstruction of the forward trajectory \cite{ijcai2019p103,pmlr-v119-zhuang20a}. Recent work has focused on including differentiable formulations of ODE systems directly in optimization, control, and reinforcement learning workflows \cite{pmlr-v283-miao25a,11128641}, motivating the need to work with differentiable formulations of dynamical systems. 

A typical calibration exercise requires the user to translate equations into code, connect the code to observation data, define parameter bounds and transformations, select suitable numerical solvers, choose estimation algorithms, and diagnose failures when the fit is poor. 

Executing all of the above steps requires expertise and effort to setup the entire estimation process. Any of these steps, if done incorrectly, can cause failure of the estimation. Some common failure modes include inappropriate scaling of parameters and objectives, an incorrect optimization setup, insufficient estimation algorithms, poor objective function construction, insufficient constraint flexibility, too broad or too narrow search ranges , and inefficient code implementations.

\subsection{Agentic AI workflows}

AI agents have recently been adopted across scientific computing for tasks such
as geometry and CAD creation \cite{khan2024text2cad,guan2026cad}, solver orchestration \cite{chen2024metaopenfoam,yue2025foamgpt}, end-to-end simulation
pipelines and post-processing \cite{xu2025cfdagent,chen2025metaopenfoam2}.
These workflows share a characteristic profile: they require substantial manual setup, proceed largely by trial and error, expose a large number of configuration knobs that materially affect the solution yet are easily overlooked, and demand expert knowledge to operate the underlying tools effectively. Crucially, their outcomes are often \emph{verifiable}- the task’s output can be checked against clear evidence, rules, tests, or ground truth rather than judged only by subjective preference. This combination of high manual overhead,
expertise requirements, and built-in verifiability makes such workflows particularly amenable to augmentation by AI agents: the agent can shoulder the routine and knowledge-intensive setup. The success of these efforts, together with the recurring need for manual intervention and domain expertise, directly motivates applying AI agents to the ODE parameter estimation problem considered here.

However, delegating complex problems to an AI agent without user oversight risks hallucination, context degradation, and
silent, incorrect assumptions, any of which can yield erroneous solutions \cite{xu2025cfdagent,park2026selfcorrecting}.  An agent left to run unsupervised through a long chain of decisions can therefore accumulate and amplify errors that a domain expert would catch immediately. A potential solution to these problems is using models with large parameter counts, large context windows, and extended reasoning capabilities. Models provided by major AI providers through cloud-based services can be used to solve such problems via effective tool calling and the ability to directly process a large number of tokens. However, this methodology comes with issues around reproducibility and auditability, where the agent may make silent assumptions on different runs of the same problem and not surface them. More importantly, using cloud-hosted models to solve problems in scientific computing comes with privacy and intellectual property concerns, and often, practitioners insist on their data, mathematical models and workflow knowledge not leaving air-gapped computing environments that are locally managed.

The workflow described in this chapter is designed around these constraints, and describes a method to orchestrate local coding models to solve ODE parameter estimation problems. The broader fitting problem is decomposed into a sequence of sub-steps, each completed by the agent and producing intermediate artifacts written to disk. The decomposition enables effective utilization of context windows, and the tool performs a series of deterministic and LLM-based evaluations at every step. The tool identifies inconsistencies in the problem specification, repairs common input pathologies, transforms user-defined model functions into JIT-compilable (just-in-time compiled) routines, and assembles a two-stage optimization pipeline
that combines a parallelized global search with gradient-based refinement. The models used in this workflow need not exceed approximately 40 billion parameters, making local deployment feasible on suitably equipped workstations  

This chapter is accompanied by two repositories: \href{https://github.com/Saakaarb/pfit-local-agent}{\texttt{pfit-local-agent}}, which implements the workflow using a locally deployed LLM, and \href{https://github.com/Saakaarb/pfit-claude/tree/main}{\texttt{pfit-remote-agent}}, which orchestrates the workflow through a set of skill files using remotely hosted LLMs. The discussion focuses on \texttt{pfit-local-agent} to address the additional constraints associated with local deployment.

\section{Methods}

\subsection{Context Assisted Agents via division of tasks}

A transformer language model defines an autoregressive distribution
$p(x_t \mid x_{<t})$ over the next token $x_t$ given all preceding tokens in its
context window, repeatedly predicting the next piece of text from
everything written so far. This conditioning is implemented by a mechanism called
self-attention: as the model processes each token, it scores how relevant every
earlier token is to the one currently being processed, using this relevance score to predict the next token. At inference time, the context guides how the model applies its learned capabilities to the current task: instructions, reference material, and examples are therefore important levers for controlling an agent.

The context window is also a finite, shared resource. For each token, the relevance scores described above are normalized so that they sum to one across all the earlier tokens (an operation called a softmax). Although attention weights are normalized, the model can assign very small weights to irrelevant tokens; increasing context length therefore does not necessarily reduce attention to relevant information. Nevertheless, loading extensive material into a single prompt increases computational demands and may make it harder for the model to select and consistently apply the instructions relevant to a particular step.

Hence, dividing a large task into sub-tasks, that determine which material is loaded into the context window precisely when that step executes, is effective: the rules the step must follow, the
structural templates its output must match, and tool calls the LLM may make. Because the model attends over the sub-task, detail belonging to other steps is kept out of the window
and out of the attention budget. This also keeps the active window dense with relevant tokens
and well within its length limit. This both sharpens output quality and lets each step be revised independently by editing its context alone.

\subsection{Search Algorithms}

\subsubsection{Population methods}

The framework ships with two population-based global optimizers, Particle Swarm
Optimization (PSO) and Differential Evolution (DE), for the coarse search that
precedes gradient-based refinement. Both are \emph{gradient-free}, tolerating
the stiff, non-smooth losses typical of ODE fitting; \emph{global}, exploring an
entire population rather than a single iterate to avoid poor local minima; and
\emph{embarrassingly parallel}, since per-candidate evaluations are independent
and dispatched in parallel. The population is given a space-filling
initialization via an optimized Latin Hypercube (LHS) design \cite{damblin2013optimizedlhs}, and both search in the
normalized $[-1,1]^d$ space so parameters spanning many orders of magnitude are
treated on equal footing. The best candidate found seeds the gradient stage.

\paragraph{Particle Swarm Optimization (PSO)}

The global search defaults to PSO\cite{kennedy1995pso}, a population-based metaheuristic in which
a swarm of $N$ candidate solutions (\emph{particles}) explores the normalized
parameter space $[-1,1]^d$. Particle $i$ carries a position
$\mathbf{x}_i \in [-1,1]^d$ and a velocity $\mathbf{v}_i$. The swarm is seeded
with a LHS design to spread the initial positions evenly, and a
star topology is used so every particle shares a single global best.

Let $f(\cdot)$ denote the (normalized) loss. Each particle tracks its own best
position $\mathbf{p}_i$ and the swarm's global best $\mathbf{g}$:
\begin{equation}
  \mathbf{p}_i \leftarrow \arg\min_{\mathbf{x} \in \{\mathbf{p}_i,\,\mathbf{x}_i\}} f(\mathbf{x}),
  \qquad
  \mathbf{g} \leftarrow \arg\min_{i} f(\mathbf{p}_i).
\end{equation}
At each iteration $k$ the velocity and position are updated by
\begin{align}
  \mathbf{v}_i^{k+1} &= w\,\mathbf{v}_i^{k}
      + c_1 \mathbf{r}_1 \odot (\mathbf{p}_i - \mathbf{x}_i^{k})
      + c_2 \mathbf{r}_2 \odot (\mathbf{g} - \mathbf{x}_i^{k}), \\
  \mathbf{x}_i^{k+1} &= \mathbf{x}_i^{k} + \mathbf{v}_i^{k+1},
\end{align}
where $\mathbf{r}_1, \mathbf{r}_2 \sim \mathcal{U}(0,1)^d$ are drawn
element-wise, $\odot$ is the Hadamard product, $w$ is the inertia weight, and
$c_1, c_2$ are the cognitive and social coefficients. The velocity is clamped
component-wise to $[-0.3, 0.3]$, and particles leaving the box are projected
back to the nearest boundary.

Relative to differential evolution, PSO tends to converge faster on easier problem landscapes: the velocity memory and shared global attractor let information
propagate through the swarm quickly, so good regions are exploited in fewer loss
evaluations.

\paragraph{Differential Evolution (DE)}

As an alternative global search, the framework provides Differential Evolution \cite{storn1997differential},
a population-based evolutionary algorithm well suited to multimodal, non-differentiable objectives. A population of $N$ candidate vectors
$\{\mathbf{x}_i\}_{i=1}^{N}$ in $[-1,1]^d$ is initialized with LHS and evolved by mutation, crossover, and selection.

For each target vector $\mathbf{x}_i$ at generation $k$, a \emph{mutant} is
formed from the current best and two distinct randomly chosen members
($r_1 \neq r_2 \neq i$):
\begin{equation}
  \mathbf{m}_i = \mathbf{x}_{\text{best}}^{k}
      + F\,\bigl(\mathbf{x}_{r_1}^{k} - \mathbf{x}_{r_2}^{k}\bigr),
\end{equation}
where the differential weight $F$ is sampled per generation
from $\mathcal{U}(0.5,\,1.0)$ to improve robustness. A trial vector
$\mathbf{u}_i$ is built by binomial crossover, mixing the mutant with the
target at a per-component rate set by the recombination constant
$CR = 0.7$:
\begin{equation}
  u_{i,j} =
  \begin{cases}
    m_{i,j}, & \text{if } \rho_j \le CR \ \text{ or } \ j = j_{\text{rand}}, \\
    x_{i,j}^{k}, & \text{otherwise},
  \end{cases}
  \qquad \rho_j \sim \mathcal{U}(0,1),
\end{equation}
where $j_{\text{rand}}$ guarantees at least one component is taken from the
mutant. Greedy selection retains whichever of the trial or target has lower
loss:
\begin{equation}
  \mathbf{x}_i^{k+1} =
  \begin{cases}
    \mathbf{u}_i, & \text{if } f(\mathbf{u}_i) \le f(\mathbf{x}_i^{k}), \\
    \mathbf{x}_i^{k}, & \text{otherwise}.
  \end{cases}
\end{equation}
Relative to PSO, DE is more robust on highly multimodal or ill-conditioned
landscapes\cite{storn1997differential}: its difference-vector mutation adapts the search scale to the
population's current spread, and the greedy, elitist selection prevents the
population from drifting away from good solutions. It is therefore the safer
choice when PSO stalls in a poor local minimum, at the cost of more loss
evaluations to converge.

\subsubsection{Gradient-based methods}

The best candidate from the global search is refined by a gradient-based
optimizer that minimizes the same normalized loss $f(\mathbf{x})$ over
$[-1,1]^d$. Gradients are obtained by \emph{automatic differentiation} (AD)
through the full discretized ODE solve, rather than by finite differences. Because AD
differentiates the discretized solver--objective map itself, it yields exact
discrete sensitivities and requires no perturbation step size---a quantity that
is notoriously hard to choose for stiff or highly nonlinear systems
\cite{baydin2018automatic,ma2021comparison}. In reverse mode, this is equivalent to a discrete adjoint approach, propagating sensitivities backward through a discretized solver to compute the loss gradient. The update is a projected step that
keeps iterates inside the search box,
\begin{equation}
  \mathbf{x}^{k+1} = \Pi_{[-1,1]^d}\!\bigl(\mathbf{x}^{k} + \Delta\mathbf{x}^{k}\bigr),
  \qquad
  \Delta\mathbf{x}^{k} = \mathcal{U}\bigl(\nabla f(\mathbf{x}^{k}),\, \mathbf{x}^{k}\bigr),
\end{equation}
where $\Pi$ is clipping onto the bounds and $\mathcal{U}$ is the optimizer's
update rule. The framework exposes both a first-order method (Adam, with an
exponentially decaying learning-rate schedule) and a second-order quasi-Newton
method (L-BFGS); the best iterate is tracked throughout and returned, and a
failed integration is caught so the search exits gracefully with the best
solution so far.

This stage is intended for \emph{fine-tuning} rather than global discovery: it
converges quickly but needs a good initial guess, which the population-based
search provides. It is resource-efficient and scales well with
dimension, since AD computes the full gradient at a cost comparable to a single
loss evaluation regardless of the number of parameters---unlike finite
differences, whose cost grows linearly with dimension.

The gradient based search is also the more fragile of the two methods, and several failure modes exist. The computed gradient inherits the accuracy of the
forward solve, so loosely converged or under-resolved integration injects noise
into $\nabla f$ and can cause the iterates to diverge; the solver tolerances
must generally be tighter for the gradient stage than for the global search to allow for a stable adjoint computation.
Excessively noisy data has a similar effect, producing a rugged loss surface
whose local gradients point away from the true optimum. Some systems are
ill-suited to differentiation: non-smooth dynamics (event triggers, clipping,
piecewise or discontinuous right-hand sides) break differentiability. Even for smooth, well-behaved
systems, convergence can stall on ill-conditioned problems
where the loss has long flat valleys (near-degenerate parameter combinations),
and a poor initial guess can leave the optimizer in the wrong basin since the
local search cannot escape it. Finally, stiff systems may drive the adaptive
solver to exceed its step budget, causing an integration failure mid-search;
the framework guards against this by tracking the best iterate and exiting
gracefully when a solve fails.

\subsection{Post-Fit Sloppiness Diagnostic}

After the fitting stage, the framework automatically attempts to characterize
how well the data constrain the fitted parameter combinations by computing the
local curvature of the loss at the saved best fit---the \emph{sloppiness}
spectrum, following the perspective of Gutenkunst \emph{et al.} ~\cite{gutenkunst2007universally}.
Let $\mathbf{u}$ denote the per-parameter axis with
$u_i = \log_{10}\theta_i$ for log-scaled parameters and $u_i = \theta_i$
otherwise. The framework maps each axis to a dimensionless optimizer coordinate,
\begin{equation}
  z_i = 2\frac{u_i-a_i}{b_i-a_i}-1,
  \qquad z_i \in [-1,1],
\end{equation}
where $a_i$ and $b_i$ are the parameter bounds expressed on the corresponding
linear or logarithmic axis. This removes dependence on the physical units used
to express the parameters, provided the values and bounds are converted
consistently. The spectrum still depends on the chosen bounds and linear/log
transformations. Writing the saved best fit as $\mathbf{z}^\star$, the framework
forms the Hessian of the configured loss in these coordinates,
\begin{equation}
  H = \nabla_{\mathbf{z}}^2 f(\mathbf{z}^\star)
  \in \mathbb{R}^{d\times d},
\end{equation}
by second-order automatic differentiation through the ODE solve, falling back
to finite differences of the first-order AD gradient when second-order
differentiation is unavailable or returns nonfinite curvature. Central
differences are used where possible, with one-sided differences near parameter
bounds. The Hessian is symmetrized, $H \leftarrow \tfrac{1}{2}(H + H^\top)$,
and diagonalized,
\begin{equation}
  H = V\Lambda V^\top, \qquad
  \Lambda = \operatorname{diag}(\lambda_1 \ge \dots \ge \lambda_d),
\end{equation}
yielding an eigenvalue spectrum $\{\lambda_k\}$ and orthonormal eigenvectors
$\{\mathbf{v}_k\}$. Each eigenvector is a \emph{direction} in normalized
parameter space, and its eigenvalue is the curvature of the loss along that
direction. Near a local minimum, large positive eigenvalues indicate
\emph{stiff} combinations, while small positive eigenvalues indicate
\emph{sloppy} combinations; eigenvalues approaching zero indicate locally flat
directions.

When the largest eigenvalue is positive, the reported verdict uses the
eigenvalues normalized by the largest, $\hat\lambda_k = \lambda_k/\lambda_1$.
In the absence of significant negative curvature, a direction is flagged as
locally weak when
\begin{equation}
  \hat\lambda_k < \varepsilon_0, \qquad \varepsilon_0 = 10^{-6}.
\end{equation}
The report also gives the span of the positive eigenvalue spectrum in decades.
Significant negative eigenvalues are reported separately as indefinite
curvature, rather than interpreted as evidence of a local minimum. This is a local, quadratic diagnostic whose interpretation as parameter
constraint is most appropriate near a converged minimum. The report therefore
includes stationarity and parameter-bound warnings. It does not establish fit
quality, global optimality, or structural identifiability.

Sloppiness concerns parameter combinations rather than individual parameters.
The reported eigenvectors indicate which parameters contribute to the local
weak directions. If repeated runs under identical estimation conditions produce
similar losses but substantially different parameter values, this can provide
additional evidence of weakly constrained combinations, although distinct
minima or incomplete convergence can also produce such variation.

\section{Procedure}
  \label{sec:procedure}

  This section describes how to configure and run an ODE parameter-estimation
  problem using the local pfit workflow. The supported starting point is a reviewed
  mathematical problem description in \texttt{inputs/user\_info.txt}, together
  with one or more CSV data files. The user is responsible for specifying the
  equations, parameter bounds, initial conditions, mappings from ODE system to observables, and mathematical
  objective. The framework then uses a language model to convert this specification into executable model code.

  The workflow is divided into five command-line stages:

  \begin{center}
  \ttfamily
  pfit new $\rightarrow$ pfit check $\rightarrow$ pfit jax $\rightarrow$
  pfit run $\rightarrow$ pfit diagnose
  \end{center}

  During \texttt{pfit new}, the configured model extracts a structured
  fitting problem from \texttt{user\_info.txt} and the declared CSV files.
  Deterministic checks control whether
  \texttt{inputs/user\_input.yaml} and \texttt{generated/user\_model.py} are
  accepted. The \texttt{pfit check} stage examines the resulting data, configuration, model declarations, and objective, and writes a report without modifying the user-provided model.

  The \texttt{pfit jax} stage uses the language model to translate the accepted
  Python model into JAX. The candidate translation is subjected to generated-code
  interface checks, numerical smoke tests, and sampled source-to-JAX comparisons. The \texttt{pfit run} stage repeats deterministic readiness checks and performs population-based exploration followed by gradient-based refinement. Each invocation creates a new preserved run directory, and post-fit sloppiness analysis is performed by default.

  Finally, \texttt{pfit diagnose} reconstructs a completed run from its saved snapshot, calculates residual and optimization diagnostics, and generates
  measured-versus-simulated and residual plots. It also summarizes the sloppiness
  results produced during fitting. The configured model may be used to interpret the computed evidence, but it does not modify the session.
  
  The separation between stages is deliberate. Each stage produces files that can be inspected before the next stage begins, making assumptions and generated code visible to the user. It also enables more efficient and accurate results from the language model, by only including necessary information in context for any step.

  The following subsections first define the software, data, and problem-specification requirements and then describe each command and its generated artifacts in detail.

\subsection{Setting up a problem}

\subsubsection{Initial software setup}
 \label{sec:init_software_setup}
  The local workflow requires two components: a Python environment for running
  \texttt{pfit}, and a locally served model exposed through Ollama. The Python package definition is stored in \texttt{pyproject.toml}, while the tested dependency set is pinned in \texttt{requirements-lock.txt}. For reproducible use, create a fresh virtual environment and install the locked environment:

  \begin{verbatim}
  python3 -m venv .venv
  source .venv/bin/activate
  python -m pip install --upgrade pip
  python -m pip install -r requirements-lock.txt
  \end{verbatim}

  It is \textbf{highly} recommended to use the software dependencies listed in the lock file, to avoid performance regressions. This installs the \texttt{pfit} command-line entry point together with the numerical and model-generation dependencies used by the workflow, including JAX, Diffrax\cite{kidger2021on}, Optax\cite{deepmind2020jax}, and the Ollama Python client. After installing Ollama, start the Ollama service and pull the selected coding model, for example:

  \begin{verbatim}
  ollama pull qwen3:30b-a3b
  \end{verbatim}

  The model used by \texttt{pfit} can be configured in \texttt{pfit.yaml}:

  \begin{verbatim}
  llm:
    model: qwen3:30b-a3b
    base_url: http://localhost:11434

  workflow:
    temperature: 0.1
    max_tokens: 12000
    max_repair_attempts: 5
  \end{verbatim}

  The local model is required for the model-assisted stages,
  \texttt{pfit new}, \texttt{pfit check}, \texttt{pfit jax}, and optional semantic diagnosis. Once a session has a validated YAML configuration and generated JAX script, \texttt{pfit run} can be repeated without calling the language model.

\subsubsection{Requirements on User Input}
 \label{sec:user-info-requirements}
  A new problem begins with the following directory structure:
  
  \begin{verbatim}
  sessions/<session-name>/
    inputs/
      user_info.txt
      <data-file>.csv
  \end{verbatim}

  where the \texttt{user\_info.txt} file is used to describe the problem setup to the language model. The user should try to make the file as complete as possible to make the estimation process error-free. Section \ref{sec:running_example} demonstrates a sample problem end to end, which shows how this input can be set up. Additionally, \texttt{pfit-local-agent} repository has several solved examples to demonstrate additional set ups.

  A few notes on the input: Equations may be written in ordinary mathematical notation, and short code snippets, and do not need to be syntactically correct code. Names must be used consistently: States, parameters, intermediate rates, observables, and forcing inputs should have distinct names that are valid Python identifiers. Names that differ by capitalization or underscores are treated as different quantities. Every symbol used in an ODE or observable must be defined as a state, parameter, forcing input, or intermediate formula. If required information is missing or contradictory, \texttt{pfit new} reports targeted \texttt{missing\_inputs} and stops. 

  Table ~\ref{tab:required_problem_definition} summarizes the
  required information in the input. 

  \small
  \begin{longtable}{@{}p{0.32\textwidth}p{0.63\textwidth}@{}}
  \caption{Information required to define a new parameter-estimation problem}
  \label{tab:required_problem_definition}\\
  \hline
  \textbf{Required information} & \textbf{Description} \\
  \hline
  \endfirsthead

  \multicolumn{2}{c}%
  {{\tablename\ \thetable{} -- continued from previous page}}\\
  \hline
  \textbf{Required information} & \textbf{Description} \\
  \hline
  \endhead

  \hline
  \endfoot

  ODE system
  &
  Every integrated state must be named and have exactly one right-hand-side
  expression. Intermediate rates and formulas referenced by the equations must
  also be defined.
  \\

  Dataset files and mappings
  &
  Every CSV column must be interpretable as time, observed output, forcing input, or uncertainty column. If a column is a forcing or uncertainty column, the user prompt should say so explicitly; otherwise the model may incorrectly treat it as
  an observed output. 
  \\

  Initial conditions
  &
  A finite initial value must be provided for every integrated state.
  For fitting a multi-dataset model, the user should explicitly say what the initial conditions are for each dataset.
  \\

  Trainable parameters
  &
  Every fitted parameter must have a name, finite lower and upper bounds, and an
  explicit choice of linear or logarithmic search coordinates.
  \\

  Fixed parameters
  &
  Every fixed parameter must have a name and finite numerical value. Values stated
  in the input specification are treated as authoritative.
  \\

  Derived observables
  &
  Any simulated quantity compared with data must be defined mathematically in
  terms of states, parameters, and previously defined formulas.
  \\

  Measured forcing inputs
  &
  Known time-dependent inputs must be associated with declared CSV columns. The
  current workflow supports finite forcing histories with linear interpolation.
  \\

  Loss definition
  &
  The objective should specify the simulated-to-measured mappings, normalization, transformations, reductions, weighting, penalties, and missing-data behavior. Alternatively, the user may explicitly request an agent-proposed objective, which must be reviewed and accepted before fitting. An explicitly supplied mathematical objective takes precedence over automated proposals. For more information see Section \ref{sec:objective_function}
  \\
  
  Numerical and output requirements
  &
  Known stiffness, required integrator behavior, tolerances, and any necessary
  output or writeout definitions should be stated when they are known to assist the model in choosing tolerances and integrators.
  \\

  \end{longtable}
  \normalsize 

  Based on the problem provided, the model automatically chooses between an explicit or implicit high order integrator, makes an estimation on maximum solver steps that should be used, calibrates the per-state tolerances that should be used, and performs solver accuracy checks after the selection of these. Units must already be consistent across the equations, data, initial conditions, parameter values, bounds, and objective. The framework does not infer units or perform unit conversions. 

\subsubsection{Requirements on the dataset files}

The framework does not modify user data files, so each CSV must already be in a
  form that the loader, validator, and ODE solver can use. The requirements are
  summarized in Table~\ref{tab:dataset_requirements}.

  \small
  \begin{longtable}{@{}p{0.34\textwidth}p{0.61\textwidth}@{}}
  \caption{Requirements for CSV data supplied to the pfit workflow}
  \label{tab:dataset_requirements}\\
  \hline
  \textbf{Requirement} & \textbf{Remarks} \\
  \hline
  \endfirsthead

  \multicolumn{2}{c}%
  {{\tablename\ \thetable{} -- continued from previous page}}\\
  \hline
  \textbf{Requirement} & \textbf{Remarks} \\
  \hline
  \endhead

  \hline
  \multicolumn{2}{r}{{Continued on next page}}\\
  \endfoot

  \hline
  \endlastfoot

  Comma-separated text
  &
  Files must use ordinary comma-separated fields. Tab-delimited and other
  separator formats are not supported by the fitting-data loader.
  \\

  Required header row
  &
  New sessions require a header row. The first header (at index 0) must be named exactly
  \texttt{time}. 
  \\

  Time is the first column
  &
  Column 0 contains the recorded time grid. Time values must be numeric, finite,
  and strictly increasing.
  \\

  Minimum size
  &
  Each CSV must contain at least two numeric data rows and at least one declared
  non-time column.
  \\

  Rectangular structure
  &
  Every row must contain the same number of fields, and the CSV width must match
  the declared column layout. A trailing comma may create an additional empty
  field and therefore cause a width mismatch.
  \\

  Numeric data cells
  &
  Data cells must contain numeric values or permitted missing-measurement values, such as \texttt{NaN}.
  \\

  Declared column meanings
  &
  Every non-time CSV column must be described in the user problem statement. The
  statement should say whether the column is a measured state, a measured derived
  observable, a measurement uncertainty, or a known forcing input. The generated
  YAML configuration records these meanings explicitly, and \texttt{pfit check}
  verifies that the generated mappings are consistent with the CSV headers and
  problem statement.
  \\

  Measurement columns
  &
  A measurement column must map unambiguously to an integrated state or derived
  observable.
  \\

  Custom loss masking
  &
  Custom and uncertainty-weighted losses are responsible for applying valid masks
  before logarithms, normalization, or division. The framework does not guarantee
  safe masking for arbitrary custom loss algebra.
  \\

  Uncertainty columns
  &
  An uncertainty column must explicitly identify the measurement column to which
  it belongs. The framework can use uncertainty for a measurement as a normalizing metric when building an RMSE loss.
  \\

  Forcing columns
  &
  A forcing column represents a known input history rather than an observation to
  be fitted. The current workflow uses linear interpolation.
  \\

  Initial conditions
  &
  Initial conditions are supplied in \texttt{user\_info.txt} and recorded in the generated configuration. They are applied at the first time point in the CSV, which need not be t = 0. Where an integrated state is also measured, the stated initial condition should be consistent with the measured value at that first time point.
  \\

  Multiple experiments
  &
  All CSVs in one fitting session must use the same ordered column names, roles,
  measurement mappings, uncertainty relationships, and units. Their row counts,
  time grids, time ranges, numerical values, and declared initial conditions may
  differ.
  \\

  Consistent units
  &
  Time, measurements, forcing inputs, initial conditions, model equations, and
  parameter values must already use compatible units. The framework neither
  infers units nor performs unit conversion.
  \\

  \end{longtable}
  \normalsize

  When generated model code is executed, the time column is passed separately as
  the integration grid. Consequently, \texttt{dataset[:, 0]} in generated Python
  or JAX code refers to the first \emph{non-time} CSV column, not to time. Users
  should keep this indexing convention in mind when reviewing the generated loss,
  forcing, and output functions.

\paragraph{Fitting multiple datasets simultaneously} 

Several experimental records may be fitted simultaneously when they describe
  the same dynamical system under different experimental conditions. The framework
  estimates one shared parameter vector and uses the same equations, fixed
  parameters, observables, loss implementation, and numerical settings for every
  record. Experiments that require different fitted parameter values should
  instead be placed in separate sessions.

  Each experiment is associated with its own CSV file and may have its own time
  grid, number of observations, initial and final times, initial-condition
  overrides, and forcing history. The problem statement must identify every file
  and state explicitly what differs between the experiments. For example:

  \begin{verbatim}
  Experiments:
  - run_low.csv:  A(0) = 0.1, B(0) = 0
  - run_mid.csv:  A(0) = 1.0, B(0) = 0
  - run_high.csv: A(0) = 10.0, B(0) = 0

  \end{verbatim}

  The scalar loss is evaluated separately for each experiment using that record's
  own data, time grid, initial conditions, and forcing values. If there are
  \(N_{\mathrm{exp}}\) experiments with scalar losses
  \(\mathcal{L}_{e}\), the fitted objective is

  \begin{equation}
  \mathcal{L}_{\mathrm{total}}
  =
  \frac{1}{N_{\mathrm{exp}}}
  \sum_{e=1}^{N_{\mathrm{exp}}}
  \lambda_{e}\mathcal{L}_{e}.
  \label{eq:equal_experiment_mean}
  \end{equation}

  By default, \(\lambda_e\)=1. This is an equal-weight mean of complete experiment losses, not a pooled loss
  over every observation. The mathematical loss should therefore define a
  suitable reduction and normalization within each experiment. Otherwise, an
  experiment whose scalar loss is numerically much larger may dominate the
  optimization despite the equal experiment weights.

  Each experiment receives its own saved trajectory:

  \begin{verbatim}
  result_solution_exp1.csv
  result_solution_exp2.csv
  ...
  \end{verbatim}

  The run summary records the loss for each experiment as well as the
  equal-experiment-mean aggregation. Run snapshots, gradient-only restarts,
  sloppiness analysis, and post-fit diagnosis all use the complete set of
  experimental records.  

\subsubsection{Objective function}
\label{sec:objective_function}

The construction of the objective is central to the quality of the obtained
  estimates. The objective \textbf{must} capture all constraints on the ODE
  solution that the user requires. Let the model solution be written as
  \(\mathbf{y}(t;\boldsymbol{\theta})\), where \(\boldsymbol{\theta}\) denotes the
  fitted parameters. A general user-specified constraint can be written as
  \begin{equation}
      f_i\!\left(t,\mathbf{y}(t;\boldsymbol{\theta}),\boldsymbol{\theta}\right)
      \leq 0,
      \label{eq:user_constraint}
  \end{equation}
  where \(f_i\) may represent a physical bound, a monotonicity requirement, a
  relationship between states, or a constraint on a derived observable. Such
  constraints can be incorporated into the objective through weighted penalty
  terms. A typical objective has the form
  \begin{equation}
      \mathcal{L}(\boldsymbol{\theta})
      =
      \mathcal{L}_{\mathrm{data}}(\boldsymbol{\theta})
      +
      \sum_{i=1}^{N_c}
      w_i \,
      \mathcal{P}_i\!\left[
      f_i\!\left(t,\mathbf{y}(t;\boldsymbol{\theta}),\boldsymbol{\theta}\right)
      \right],
      \label{eq:penalized_objective}
  \end{equation}
  where \(\mathcal{L}_{\mathrm{data}}\) measures agreement between the simulated
  outputs and the data, \(w_i\) is the weight assigned to the \(i\)-th constraint,
  and \(\mathcal{P}_i\) is a nonnegative penalty function. The penalty function should be smooth to allow stable gradient computation.

  This requires some user expertise and insight, as an underconstrained problem
  may allow unanticipated or undesired dynamics, while an overconstrained problem
  may prevent the optimizer from finding a feasible solution during the global
  optimization stage. The user may choose to allow the agent to formulate the loss
  from the problem specification. This can work for simpler problems, but for more
  complex problems it is advised that the user specify how the loss is constructed
  from the ODE solution manually. Broad best practices for constructing the
  objective function are listed in Table~\ref{tab:objective_function_guidelines}.

\begin{table}[h]

\caption{Best practices for building an objective function}
\label{tab:objective_function_guidelines}
\centering
\small
\begin{tabular}{@{}p{0.38\textwidth}p{0.57\textwidth}@{}}
\hline
\textbf{Best Practice} & \textbf{Remarks} \\
\hline
Every term of the overall objective should be of similar scale in magnitude (preferably close to unity)
  & It is recommended to divide each objective term by a scale belonging to its observable, like its own peak, value range, or measurement uncertainty. This enables more uniform estimation of parameters, and a relative importance of terms can be controlled by setting channel weights. Several assumptions in the framework are built around this scale being close to unity.\\
Weighting of samples within a trajectory
  & Some trajectories to fit to might only carry relevant dynamics in small subsets of the overall time history, but by default each sample in time history may be weighted equally. This can lead to poor estimates, and the loss formulation can be weighted unequally in time to mitigate this  \\

Scaling of value \emph{within} a loss term
  & In some instances, the data being estimated can span several orders of magnitude within a single channel and trajectory. Users are advised to use logarithmic scalings to allow for uniform contributions to the objective in such cases \\  
Differentiability of constraints
    & To allow gradient-based refinement, constraint penalties should be written in
    smooth or nearly smooth forms whenever possible. When a threshold-like behavior is needed, users should prefer a smooth approximation, such as a steep sigmoid or soft transition. \\

\hline
\end{tabular}
\end{table}

\subsection{Running example}
\label{sec:running_example}
\subsubsection{Starting a new session}
 \label{sec:starting_a_new_session}
After preparing the input statement and data files, the session directory should
  contain only the user-supplied files:

  \begin{verbatim}
  sessions/<session-name>/
    inputs/
      user_info.txt
      <data-file>.csv
      ...
  \end{verbatim}

  Create the session with:

  \begin{verbatim}
  pfit new sessions/<session-name>
  \end{verbatim}

  The \texttt{pfit new} command is the first model-assisted stage of the workflow.
  It reads the natural-language problem statement in \texttt{user\_info.txt}
  together with the headers and numerical contents of the CSV files. The configured
  local model is then asked to extract the structured fitting problem: the
  experiments to include, trainable and fixed parameters, integrated state
  variables, initial conditions, right-hand-side equations, derived observables,
  measured outputs, forcing inputs, uncertainty columns, and the intended loss
  definition.

  The framework renders the extracted information into a session
  configuration and readable model file, then applies deterministic checks before
  writing the accepted artifacts. These checks verify that required fields are
  present, CSV headers are accounted for, parameter bounds and initial conditions
  are finite, state and parameter names are consistent, forcing and uncertainty
  columns have valid roles, and the generated model exposes the expected Python
  interfaces.

  When successful, \texttt{pfit new} writes the initial generated artifacts:

  \begin{verbatim}
  sessions/<session-name>/
    inputs/
      user_input.yaml
    generated/
      user_model.py
      agent_logs/
  \end{verbatim}

  The file \texttt{inputs/user\_input.yaml} contains the structured configuration
  used by later stages, including experiment definitions, data-column mappings,
  parameter bounds, initial conditions, solver settings, and output settings. The
  file \texttt{generated/user\_model.py} contains a readable NumPy implementation
  of the ODE right-hand side, derived observables, loss calculation, and output
  writeout. The \texttt{generated/agent\_logs/} directory stores the local-model
  requests and responses used during construction.

  If the problem statement is incomplete or internally inconsistent,
  \texttt{pfit new} stops before accepting generated files and reports targeted
  \texttt{missing\_inputs}. These messages indicate what the user should clarify
  in the input statement or data files, such as missing parameter bounds, ambiguous
  CSV column meanings, absent initial conditions, or an undefined observable.
  After updating the user-supplied inputs, the session can be regenerated with:

  \begin{verbatim}
  pfit new sessions/<session-name> --overwrite
  \end{verbatim}

\paragraph{Two state drug response model}
The tutorial uses a two-state drug-response model fitted simultaneously to pulse
and washout experiments. Both CSV files use the same columns,

\begin{verbatim}
time,dose_rate,signal
\end{verbatim}

but have different forcing histories and initial conditions. The corresponding
\texttt{user\_info.txt} is:

\begin{tcblisting}{
  listing only,
  breakable,
  colback=gray!6,
  colframe=gray!35,
  boxrule=0.4pt,
  arc=1mm,
  left=2mm,
  right=2mm,
  top=1mm,
  bottom=1mm
}
Problem:
Fit pulse.csv and washout.csv using one shared parameter set.

Experiments and initial conditions:
pulse.csv:   D(0) = 0, R(0) = 0
washout.csv: D(0) = 0, R(0) = 0.3

CSV columns:
time: time in minutes
dose_rate: known forcing input in concentration per minute;
           use linear interpolation
signal: measured value of signal_model in arbitrary units

Trainable parameters:
k_elim:   bounds [0.01, 10], logarithmic
k_on:     bounds [0.001, 10], logarithmic
k_off:    bounds [0.001, 10], logarithmic
baseline: bounds [50, 150], linear
gain:     bounds [1, 200], linear

Fixed parameters:
None.

ODE system:
dD/dt = dose_rate - k_elim*D
dR/dt = k_on*D*(1 - R) - k_off*R

Observable:
signal_model = baseline + gain*R
CSV column signal measures signal_model.

Loss:
For each experiment e, use only finite signal measurements.

scale_e = max(abs(finite signal values in experiment e))

L_e = sqrt(mean(
    ((signal_model_e - signal_e)/scale_e)^2
))

The total loss is the equal-weight mean:
L = (L_pulse + L_washout)/2.

Outputs:
Write time, D, R, measured signal, and signal_model.
\end{tcblisting}

This statement supplies the scientific decisions required by the workflow. The
local model is not expected to choose missing equations, parameter bounds,
initial conditions, data mappings, units, or loss terms. The measured CSV column
\texttt{signal} is valid here because the statement explicitly maps it to the
derived observable \texttt{signal\_model}. In general, measured columns should
map unambiguously to declared integrated states or derived observables, and
forcing or uncertainty columns should be identified by role.

\subsubsection{Checking the session}
\label{sec:checking_the_session}

After reviewing the files produced by \texttt{pfit new}, check the session with:

\begin{verbatim}
pfit check sessions/<session-name>
\end{verbatim}

This command reads the generated YAML, source model, and every declared CSV
file. The framework first performs deterministic checks by parsing these files
and comparing their declarations. It verifies that the data dimensions and
headers match the configured columns, parameter values and bounds are valid,
initial conditions are complete, model names are consistent, and the source
model uses the expected interfaces and column indices.

The loss is also compared with the generated implementation. For example, if
\texttt{user\_info.txt} requests an RMSE, logarithmic residuals, uncertainty
weighting, finite-value masking, a particular normalization, or a specific
experiment aggregation, the generated loss must contain the corresponding
operation. However, if an explicit user-supplied loss is provided, it overrules any automated loss function generation. The check surfaces potential scaling issues as warnings, which are crucial to fit quality.

The deterministic checks also look for several common failure modes: non-smooth
or branch-like dynamics that may be difficult to differentiate, measured
columns with large dynamic range, and loss implementations that appear to
violate any explicit loss instructions in \texttt{user\_info.txt}.

For the running example, the framework verifies that both CSV files use the
same ordered columns; that \texttt{dose\_rate} is declared as a finite forcing
input; that \texttt{signal} maps to the derived observable
\texttt{signal\_model}; and that the two experiments retain their specified
initial conditions. It also checks that the generated loss uses finite signal
measurements, applies the stated per-experiment normalization, calculates an
RMSE for each experiment, and preserves the equal-experiment aggregation.

The report is written to:

\begin{verbatim}
sessions/<session-name>/generated/user_input_check.txt
\end{verbatim}

After performing a check, three categories of errors are surfaced. Critical errors must be corrected before continuing. Warnings identify
limitations or potentially fragile assumptions, while recommendations identify choices that the user may wish to reconsider. By default, the configured local model may also review the session description, generated source, and deterministic evidence for semantic inconsistencies.

It should be noted that passing the checks confirms that the session satisfies the implemented structural, numerical, and consistency rules for that stage. It does not establish scientific correctness or model optimality for the data.

\subsubsection{Generating the JAX model}
\label{sec:generating_the_jax_model}
Once the session has passed the checks described above, generate the executable
JAX model by running:

\begin{verbatim}
pfit jax sessions/<session-name>
\end{verbatim}

This command converts the readable NumPy model in
\texttt{generated/user\_model.py} into the differentiable JAX implementation
used for parameter estimation. The workflow breaks the code generation into smaller steps and completes translation. The resulting script is written to:

\begin{verbatim}
sessions/<session-name>/generated/generated_script.py
\end{verbatim}

A simplified excerpt of the transformed JAX system for this example is:

\begin{tcblisting}{
  listing only,
  breakable,
  colback=gray!6,
  colframe=gray!35,
  boxrule=0.4pt,
  arc=1mm,
  left=2mm,
  right=2mm,
  top=1mm,
  bottom=1mm
}
@jax.jit
def user_defined_system(t, y, other_args):
    constants = other_args["constants"]
    trainable_variables = other_args["trainable_variables"]
    dataset = constants["dataset"]
    t_eval = constants["t_eval"]

    k_elim = unscale_value(trainable_variables[0], 0.01, 10.0, True)
    k_on = unscale_value(trainable_variables[1], 0.001, 10.0, True)
    k_off = unscale_value(trainable_variables[2], 0.001, 10.0, True)
    baseline = unscale_value(trainable_variables[3], 50.0, 150.0, False)
    gain = unscale_value(trainable_variables[4], 1.0, 200.0, False)

    D = y[0]
    R = y[1]
    dose_rate = jnp.interp(t, t_eval, dataset[:, 0])

    dDdt = dose_rate - k_elim * D
    dRdt = k_on * D * (1.0 - R) - k_off * R
    return jnp.array([dDdt, dRdt])

@jax.jit
def _compute_loss_value(constants, trainable_variables,
                          solution_time, solution):
      dataset = constants["dataset"]
      baseline = unscale_value(trainable_variables[3], 50.0, 150.0, False)
      gain = unscale_value(trainable_variables[4], 1.0, 200.0, False)

      R = solution[:, 1]
      signal_model = baseline + gain * R
      signal = dataset[:, 1]

      # Fixed-size masking is compatible with JIT.
      mask = jnp.isfinite(signal)
      observed = jnp.where(mask, signal, 0.0)
      predicted = jnp.where(mask, signal_model, 0.0)
      count = jnp.sum(mask)

      scale = jnp.max(jnp.abs(observed))
      safe_scale = jnp.where(scale > 0, scale, 1.0)
      residual = (predicted - observed) / safe_scale
      loss = jnp.sqrt(jnp.sum(residual**2) / jnp.maximum(count, 1))

      # Reject empty observations or undefined peak normalization.
      return jnp.where((count > 0) & (scale > 0), loss, jnp.nan)
\end{tcblisting}

This excerpt omits the surrounding scaffolding, but it shows the main
transformation: parameters are mapped from optimizer coordinates to physical
values, the forcing input is interpolated from the non-time dataset columns, and the functions are written in a \texttt{jit-compilable} method. 

After generation, the framework first checks that the script provides the required runtime interface for integration, loss evaluation, and output writing. It then imports the script, runs the model for each dataset, verifies that trajectories and losses are
finite, and evaluates the generated code on some sampled parameter points. The JAX implementation is also compared with the readable NumPy source model at
sampled states and parameter values. The comparison checks differential-equation values, observable predictions, output columns, and losses where applicable.

\paragraph{Automated setup tuning}

The framework adaptively tunes some heuristics of the ODE solver at this stage. For instance, if too few sampled parameter points integrate successfully because the solver
reaches its step limit, the workflow may increase \texttt{max\_steps} up to the
configured recovery ceiling and rerun the same validation. If recovery succeeds, the accepted step budget is written back to the YAML configuration. Similarly, proposed state-scaled tolerances or population-stage relaxations are accepted only when they preserve finite losses on a meaningful fraction of sampled candidates. Solver-accuracy checks then compare selected feasible probes with tighter reference solves, until the tolerance gets too tight for successful integration.

\paragraph{Repair of incorrect translation}
If a translation comparison or interface check fails, the local model does a targeted repair on the fragment responsible for the failure. The number of repair attempts is limited. Before fitting, the user should inspect \texttt{generated\_script.py} and the generated reports and confirm that the equations, parameter names, observable definitions, loss function, output columns etc. match the intended workflow

Finally, re-verify that the complete session is ready to run:

\begin{verbatim}
pfit check sessions/<session-name> --ready
\end{verbatim}

\subsubsection{Running the fit}
\label{sec:running_the_fit}
After the JAX model has been generated and the session passes readiness checks,
run parameter estimation with:

\begin{verbatim}
pfit run sessions/<session-name>
\end{verbatim}

\textbf{This stage does not require the local LLM model, and can be run repeatedly on a CPU-only device}. It uses the accepted YAML
configuration, CSV files, and \texttt{generated\_script.py}. Before fitting
starts, the same deterministic readiness checks are repeated to ensure that the
generated script is current.

Each invocation creates a new preserved run directory under:

\begin{verbatim}
sessions/<session-name>/outputs/
\end{verbatim}

The default run performs population-based global exploration followed by
gradient-based refinement. The population stage searches the bounded parameter
space in the normalized coordinates defined by the YAML configuration. The
gradient stage starts from the best population result and refines it using the
differentiable JAX loss.

For an additional refinement starting from a previous run, use:

\begin{verbatim}
pfit run sessions/<session-name> gradient-only --from-run <run-id>
\end{verbatim}

By default, the run also performs post-fit sloppiness analysis at the final parameter estimate. A completed run indicates that the configured optimization procedure finished and wrote its artifacts.

\subsection{Diagnosing a completed fit}

After parameter estimation has finished, examine the result with:

\begin{verbatim}
pfit diagnose sessions/<session-name> <run-id>
\end{verbatim}

The run identifier is the name of a directory under
\texttt{sessions/<session-name>/outputs/}. If it is omitted, the most recent
completed run is selected.

Diagnosis is based on the saved run snapshot: the YAML configuration, data
files, generated JAX script, generated source model, fitted parameters, and
runtime provenance copied when the fit was run. It does not depend on the
current working files in the session. A previous result can therefore still be
diagnosed after the working YAML, data, or generated code have been changed.

For each dataset, the command recreates the fitted trajectory and compares it
with the measurements. In the running example, separate plots are produced for
the pulse and washout experiments. These show the measured signal together with
the fitted \texttt{signal\_model}. Residual values are still written to CSV
files and summarized in the diagnosis report, but they are not drawn as a
separate subplot in \texttt{fit\_expN.png}.

The diagnosis reports residual summaries, including root-mean-square error,
mean absolute error, bias, largest absolute residual, and simple time-pattern
evidence. Missing measurements are excluded from residual statistics, but
nonfinite model predictions are treated as simulation failures rather than
missing data. The report also checks whether fitted parameters are close to
their bounds, whether optimizer logs suggest that improvement was still
occurring, and whether the run used a full fit or a gradient-only restart.

If post-fit sloppiness analysis was performed during \texttt{pfit run}, the
diagnosis includes its status and summary. The diagnosis command does not
normally repeat the sloppiness calculation.

The main files produced are:

\begin{verbatim}
fit_diagnosis.txt
scientific_diagnosis.json
fit_expN.png
residual_expN_channelM.csv
\end{verbatim}

Here, \texttt{N} identifies the experiment and \texttt{M} identifies the
measured output channel. The text report is the most convenient place to begin,
while the plots and residual CSV files allow the user to inspect individual
datasets and output channels.

\begin{figure}[t]
  \centering
  \begin{subfigure}[t]{0.48\linewidth}
    \centering
    \includegraphics[width=\linewidth]{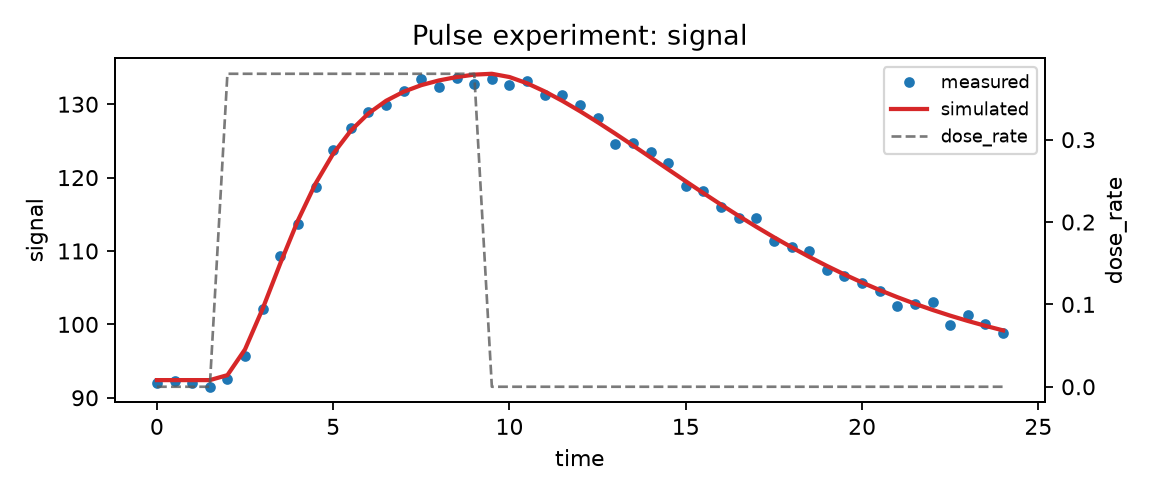}
    \caption{Pulse experiment.}
  \end{subfigure}
  \hfill
  \begin{subfigure}[t]{0.48\linewidth}
    \centering
    \includegraphics[width=\linewidth]{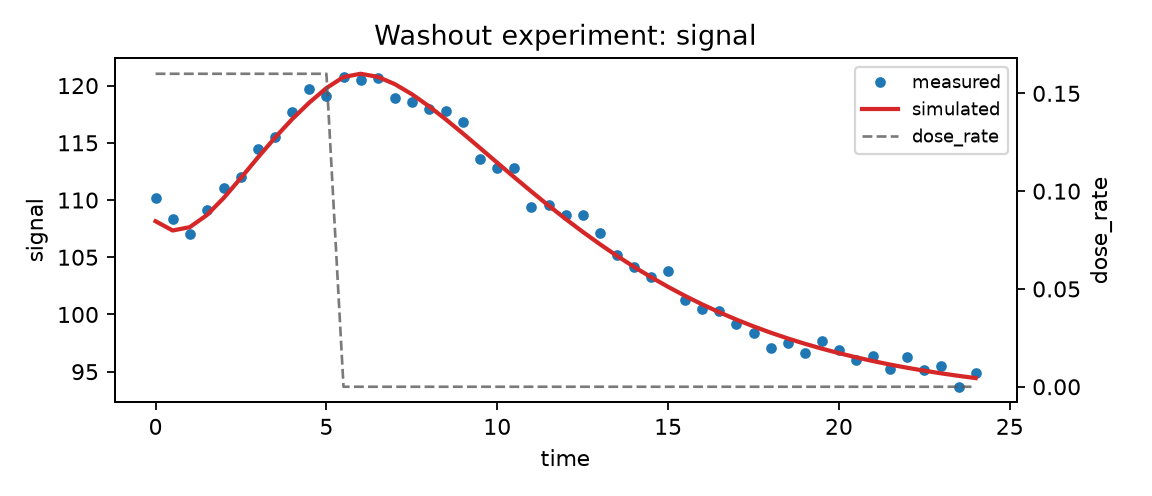}
    \caption{Washout experiment.}
  \end{subfigure}

  \vspace{0.6em}
  \begin{subfigure}[t]{0.42\linewidth}
    \centering
    \includegraphics[width=\linewidth]{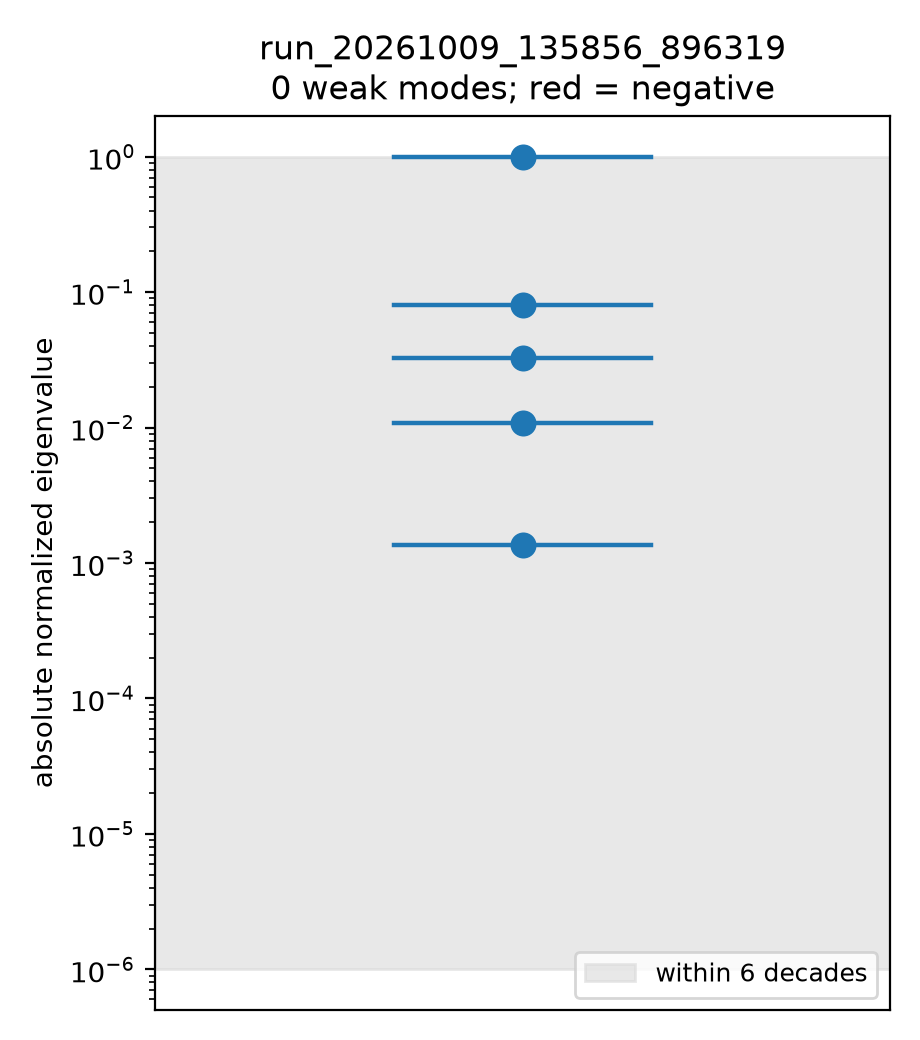}
    \caption{Post-fit sloppiness spectrum.}
  \end{subfigure}
  \caption{Example diagnostic outputs for the pulse/washout running example.}
  \label{fig:pfit-drug-response-diagnosis}
\end{figure}

For the running pulse/washout example, representative diagnostic plots are shown
in Figure~\ref{fig:pfit-drug-response-diagnosis}. These figures were generated
from synthetic data created from the same model, with small added measurement
noise, followed by the ordinary \texttt{pfit run} and \texttt{pfit diagnose}
workflow.

A representative excerpt from \texttt{fit\_diagnosis.txt} for this example is:

\begin{tcblisting}{
  listing only,
  breakable,
  colback=gray!6,
  colframe=gray!35,
  boxrule=0.4pt,
  arc=1mm,
  left=2mm,
  right=2mm,
  top=1mm,
  bottom=1mm
}
Fit diagnosis

Artifacts:
- final_design_point.csv: present
- result_solution_exp1.csv: present
- result_solution_exp2.csv: present
- de_fitting.log: present
- NODE_fitting.log: present
- fitting_error.txt: missing

Optimizer summary:
- DE: 120 iterations, first loss 8.0936E-02,
      best loss 5.7755E-03, final loss 5.7755E-03
- NODE: 21 iterations, first loss 5.7754E-03,
        best loss 5.7753E-03, final loss 5.7753E-03

Experiments: 2
Aggregation: equal_experiment_mean
- Experiment 1: pulse.csv; loss=0.005219267638936577;
  output=result_solution_exp1.csv
- Experiment 2: washout.csv; loss=0.006331386422963499;
  output=result_solution_exp2.csv

Scientific diagnosis:
- Status: ok
- Refinement stopped at its iteration budget.
- [BUDGET] Refinement stopped at its iteration budget.
  Next: Assess the final gradient and trajectory plots before
  deciding whether a gradient-only restart is worthwhile.
- Figure: fit_exp1.png
- Figure: fit_exp2.png
- Detailed evidence: scientific_diagnosis.json
\end{tcblisting}

When sloppiness analysis is enabled, a representative
\texttt{sloppiness\_report.txt} excerpt is:

\begin{tcblisting}{
  listing only,
  breakable,
  colback=gray!6,
  colframe=gray!35,
  boxrule=0.4pt,
  arc=1mm,
  left=2mm,
  right=2mm,
  top=1mm,
  bottom=1mm
}
Post-fit sloppiness
Status: ok
Method: finite-difference
Coordinates: normalized(log10(k_elim)), normalized(log10(k_on)),
             normalized(log10(k_off)), normalized(baseline),
             normalized(gain)
Verdict: no weak curvature directions at the specified relative floor
Weak modes: 0
Negative modes: 0
Normalized gradient infinity norm: 4.52654e-05
Eigenvalues (largest to smallest):
  mode 1: 9.40074106e+01
  mode 2: 7.52673682e+00
  mode 3: 3.07066809e+00
  mode 4: 1.01790654e+00
  mode 5: 1.27045465e-01
Positive spectrum span: 2.8692 decades
STIFFEST direction: k_off(-0.53), baseline(+0.51),
                   gain(+0.41), k_on(+0.39)
SLOPPIEST direction: k_elim(+0.77), k_off(-0.62),
                    baseline(-0.14), k_on(+0.07)
Warning: The normalized-coordinate gradient exceeds 1e-5;
this point is not certified stationary.
Warning: Local curvature only: not proof of structural/global
identifiability or convergence.
Warning: Curvature uses dimensionless optimizer coordinates
in [-1, 1]; the spectrum depends on parameter bounds and
linear/log transforms.
\end{tcblisting}

By default, the framework may ask the configured local language model to
summarize the computed evidence and suggest next steps. This interpretation is
advisory: the model receives the numerical diagnosis, but it does not modify the
session or the run. To produce only the deterministic report without contacting
the local model, use:

\begin{verbatim}
pfit diagnose sessions/<session-name> <run-id> --deterministic-only
\end{verbatim}

When a closer examination of local convergence is needed, the optional
\texttt{--probe-gradients} flag compares automatic derivatives with bounded
finite-difference estimates for a small selection of parameter axes:

\begin{verbatim}
pfit diagnose sessions/<session-name> <run-id> --probe-gradients
\end{verbatim}

This check can be computationally expensive and is not required for routine
diagnosis.

The diagnosis should be treated as evidence for scientific review, not as a
pass--fail judgment. A low loss does not prove that the fitted parameters are
identifiable, that the global optimum has been found, or that the model is
scientifically appropriate. The user should inspect the trajectories, residuals,
parameter locations, optimizer history, and sloppiness summary before deciding
whether the fit answers the intended scientific question.

\section{Model Performance}

This section discusses the tradeoffs of using different LLMs for the process. The workflow is evaluated on a suite of ODE parameter-estimation tasks using three locally served Ollama models. The focus is on problems with experimental data, and several examples from biological sciences are taken directly from an external, community-curated benchmark collection (the PEtab benchmark suite~\cite{hass2019benchmark}).

The framework is evaluated using three different LLM models, with the comparison run on an NVIDIA H100 80GB HBM3 GPU.  The Ollama request configuration used a 32k token context window, temperature 0.1, and a 12k token generation budget. The temperature value, which controls randomness, is kept deliberately low for this application which requires high reproducibility. Table \ref{tab:model-completion-rates} discusses memory approximate memory consumed for each model, along with their success rates on tested problems. The tests show that using newer and larger models can have a higher success rate, but they can also be slower or have more hardware requirements. For example, Qwen2.5-Coder 14B can also be run for these problems on an Apple M4.

\begin{table}[t]
\centering
\caption{Memory requirements, average completion time (all 5 stages, passed cases only) and completion rates for attempted ODE estimation problems}
\label{tab:model-completion-rates}
\begin{tabular}{@{}lrrrrr@{}}
\toprule
Model & Memory Req. (GB) & Mean run time (s) & Pass & Fail & Success Rate\\
\midrule
Qwen2.5-Coder 32B (A) & 19.9 & 110.0 & 15 & 6 & 71.4\% \\
Qwen2.5-Coder 14B (B) & 9.0 & 93.8 & 13 & 8 & 61.9\% \\
Qwen3.8-27B (C) & 17.7 & 235.6 & 19 & 2 & 90.5\% \\
\bottomrule
\end{tabular}
\end{table}

Table~\ref{tab:model-completion} summarizes the observed completion behavior, where identical tasks were provided to 3 different models. Here, a task is counted as passed when all five stages ran to completion and produced a fitted result. Qwen3.8-27B completed 19 of 21 tasks (90.5\%), compared with 15 of 21 for the 32B baseline (71.4\%) and 13 of 21 for the 14B baseline (61.9\%).  The Qwen3.8 model was slower in wall-clock time, partly because it is a thinking model and its reasoning tokens are included in the generation budget, but it was substantially more reliable on the larger biochemical and multi-experiment cases. The Qwen2.5-Coder 14B model was fast and often
adequate for small canonical systems, but it failed on several larger
multi-experiment or biochemical prompts during \texttt{pfit new}.

The retained examples span a deliberately broad range of ODE fitting tasks. They include low-dimensional textbook dynamics, stiff chemical kinetics, oscillatory systems, biochemical signaling networks, pharmacokinetics, hysteresis, measured forcing functions, derived observables, and multi-experiment shared-parameter fits. The diversity of examples demonstrates that the workflow works well in qualitatively different modeling regimes.

\begin{longtable}{@{}p{0.27\linewidth}ccc p{0.30\linewidth}@{}}
\caption{End-to-end local workflow completion by models across a diversity of ODE tasks. A check mark indicates that all five pfit stages completed without error and produced a fitted result. Models A,B,C are as described in Table \ref{tab:model-completion-rates}.}
\label{tab:model-completion}\\
\toprule
Case & Model A & Model B & Model C & Source or model family \\
\midrule
\endfirsthead
\toprule
Case & 32B & 14B & Qwen3.8-27B & Source or model family \\
\midrule
\endhead
ARC fitting & \pass & \pass & \pass & Accelerating-rate calorimetry battery example \cite{Bhatnagar_2025} \\
Boehm STAT5 & \pass & \pass & \pass & STAT5 signaling model \cite{boehm2014stat5} \\
Lotka--Volterra & \pass & \pass & \pass & Predator--prey equations \cite{lotka1925elements,volterra1926fluctuations} \\
MAPK cascade & \pass & \pass & \pass & Kholodenko MAPK cascade \cite{kholodenko2000negative} \\
Oregonator & \pass & \pass & \pass & Field--Noyes/Tyson BZ reaction model \cite{field1974oscillations,tyson1985singular} \\
Piezo Bouc--Wen & \pass & \fail & \pass & Bouc--Wen hysteresis family \cite{bouc1971modele,wen1976method} \\
Robertson & \pass & \pass & \pass & Stiff chemical kinetics model \cite{robertson1966solution} \\
Sliding basepoint & \fail & \pass & \pass & Frictional sliding-basepoint example \cite{mca29010009} \\
Theophylline & \pass & \pass & \pass & One-compartment pharmacokinetic example \cite{pinheiro2000mixed} \\
Van der Pol & \pass & \pass & \pass & Nonlinear oscillator \cite{vanderpol1926relaxation} \\
Cascaded tanks & \pass & \fail & \pass & Two-tank process-control example \cite{schoukens2016cascaded} \\
Sneyd IP$_3$ receptor & \fail & \pass & \pass & IP$_3$ receptor model \cite{sneyd2002dynamic} \\
Beer indigoidine & \pass & \pass & \pass & Indigoidine production dataset \cite{beer2014indigoidine} \\
Raia IL-13 & \pass & \fail & \pass & IL-13/JAK2/STAT5 lymphoma model \cite{raia2011il13} \\
\bottomrule
\end{longtable}

\begin{figure}[t]
\centering
\begin{subfigure}{0.48\linewidth}
  \centering
  \includegraphics[width=\linewidth]{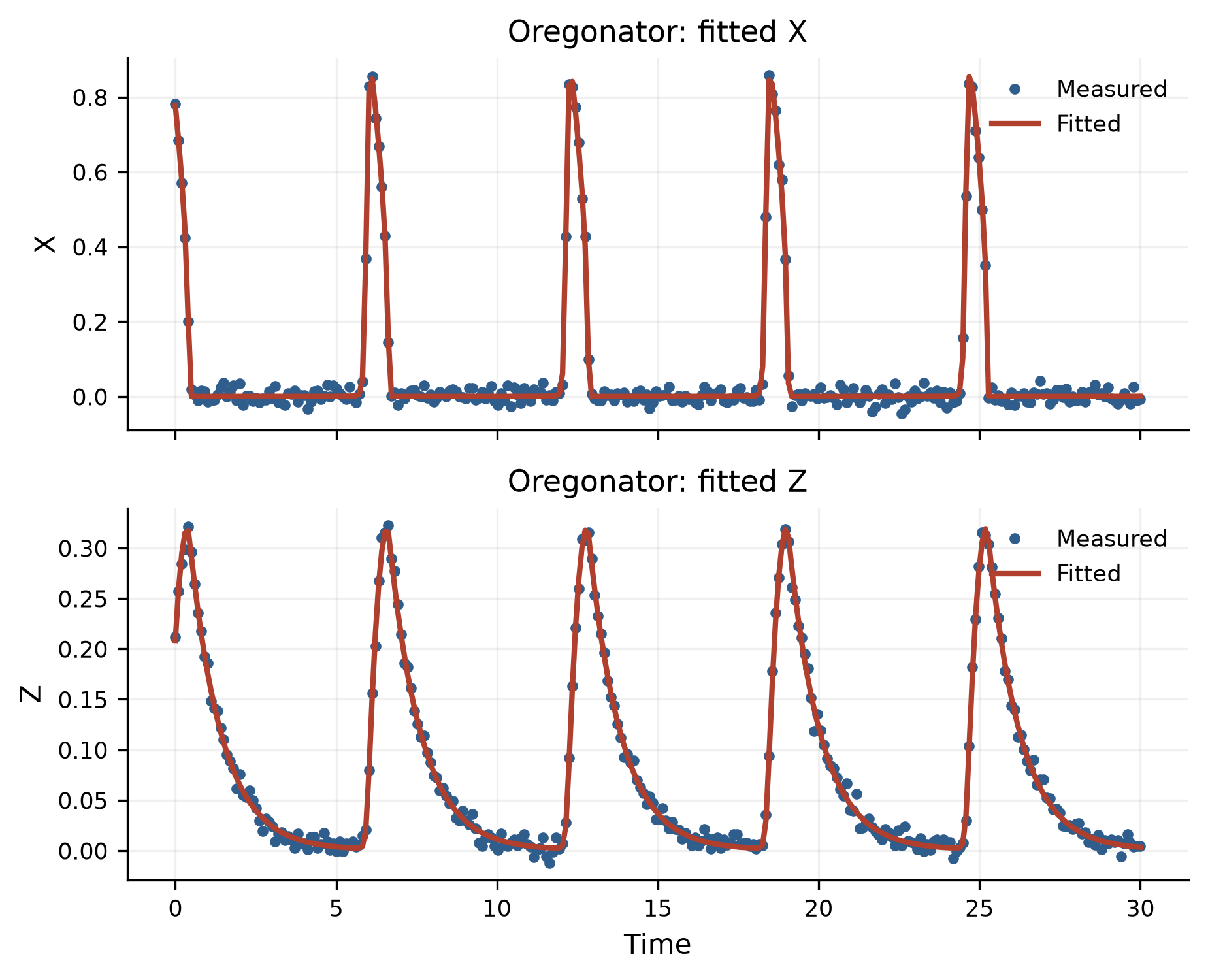}
  \caption{Oregonator oscillatory chemistry.}
  \label{fig:example_fit_oregonator}
\end{subfigure}
\hfill
\begin{subfigure}{0.48\linewidth}
  \centering
  \includegraphics[width=\linewidth]{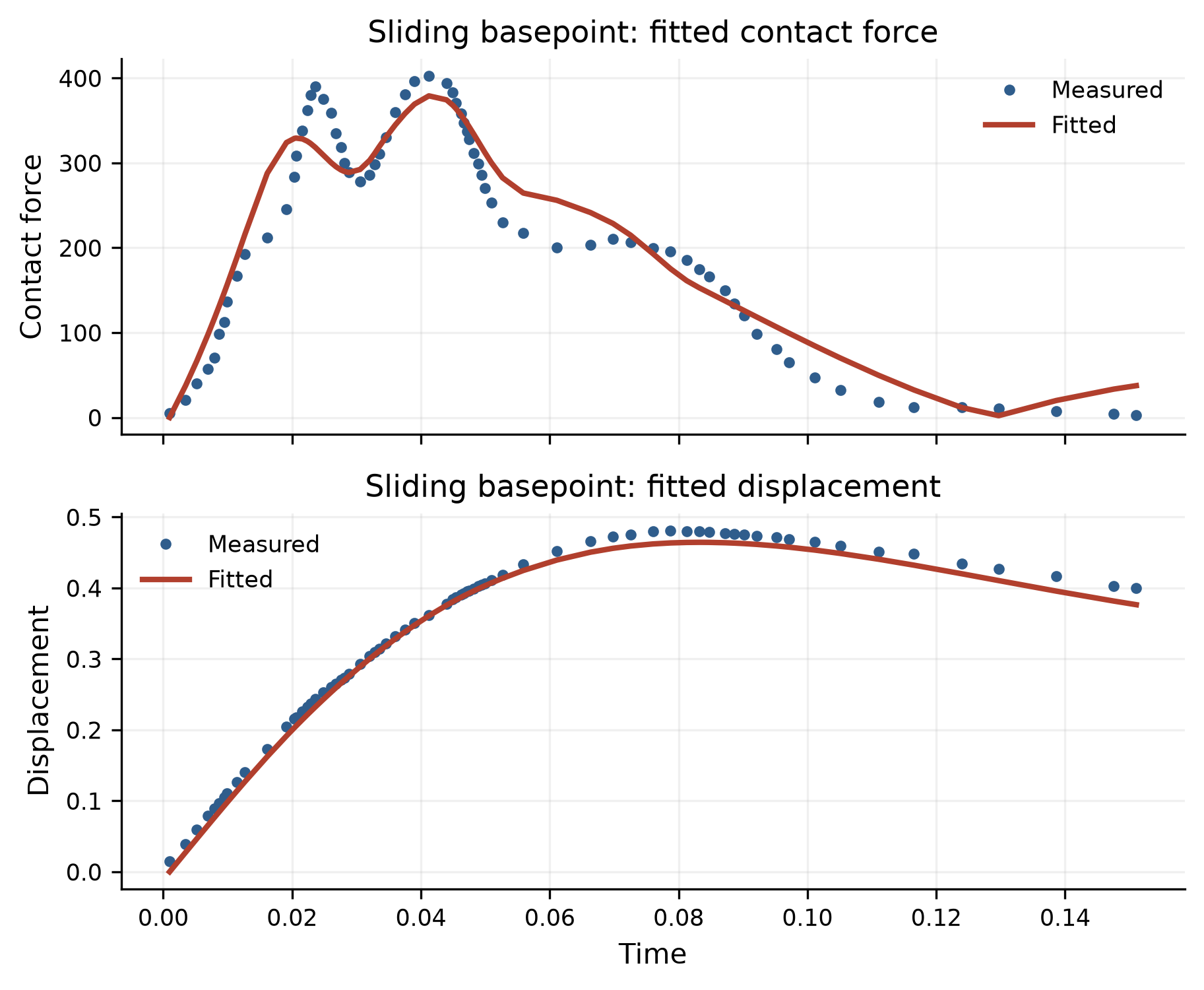}
  \caption{Sliding-basepoint friction model.}
  \label{fig:example_fit_sliding_basepoint}
\end{subfigure}

\vspace{0.5em}
\begin{subfigure}{0.48\linewidth}
  \centering
  \includegraphics[width=\linewidth]{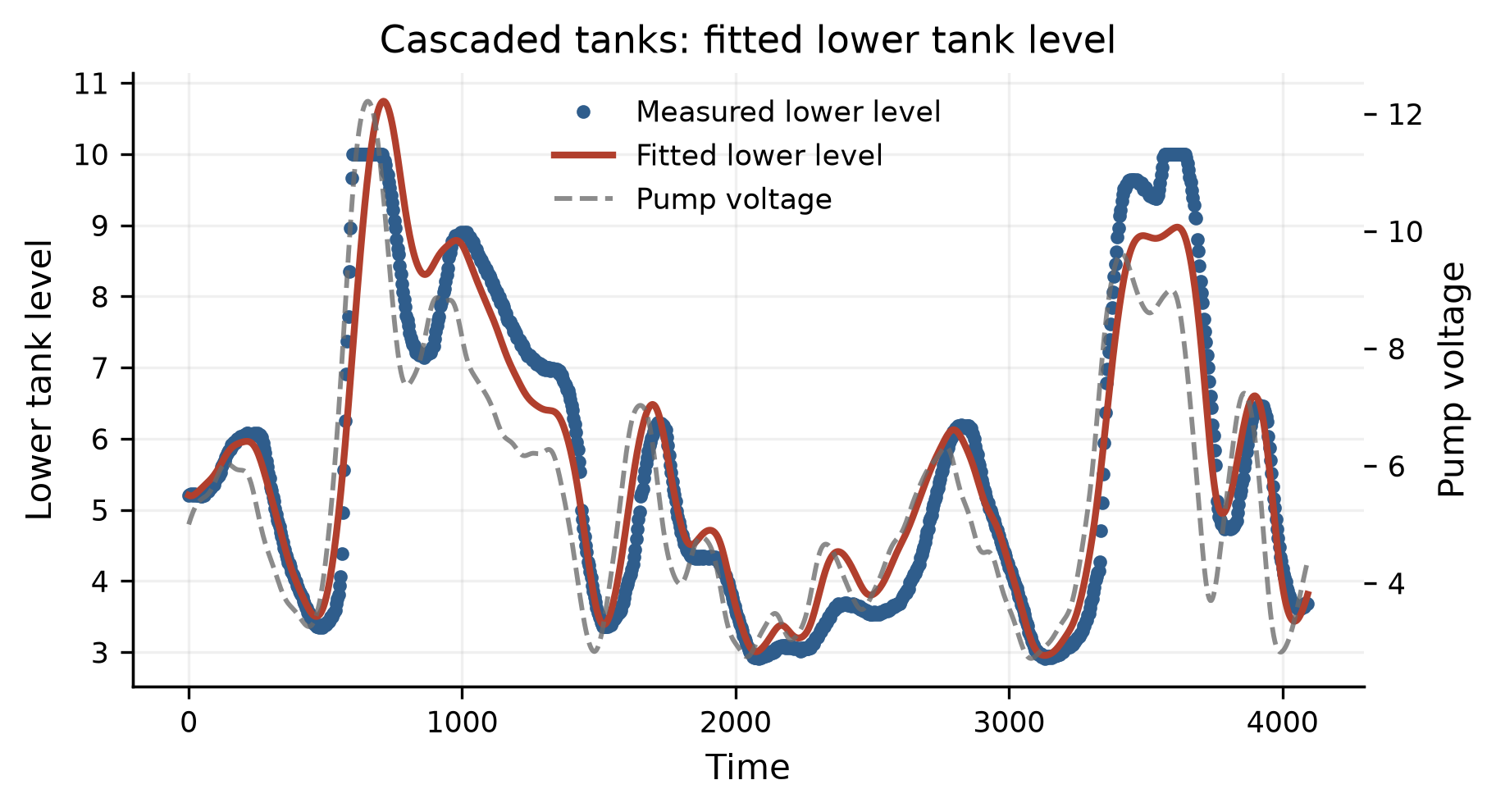}
  \caption{Cascaded tanks with measured forcing.}
  \label{fig:example_fit_cascaded_tanks}
\end{subfigure}
\hfill
\caption{Representative fits across three modeling settings: (a) stiff oscillatory dynamics in the Oregonator model, (b) a nonsmooth sliding-basepoint friction model with derived observables, and (c) cascaded-tank dynamics driven by a measured forcing input. }
\label{fig:representative-fits}
\end{figure}

 Figure \ref{fig:example_fit_oregonator} shows the Oregonator example, which tests a stiff oscillatory chemical system with only a subset of states observed. Figure \ref{fig:example_fit_sliding_basepoint} demonstrates the sliding-basepoint case, which tests a nonsmooth frictional model with derived force and displacement observables. Finally, Figure \ref{fig:example_fit_cascaded_tanks} shows the cascaded-tanks example which tests a measured forcing input. These are examples of nontrivial estimation tasks

The completed examples are included in the repository under \texttt{sessions/} at the linked repository. Each curated session contains the user prompt and data under \texttt{inputs/}, the generated model artifacts under \texttt{generated/}, and at least one completed fitting run under \texttt{outputs/}.  The \texttt{outputs/} directory for each retained case also contains one or more \texttt{fit\_exp*.png} figures showing the fitted trajectory against the data. These sessions are intended to serve both as examples for users and as reproducibility evidence for the local-agent workflow.

It should be noted that successful execution of the workflow is not the same thing as proving that the proposed ODE is a perfect description of the system.  In several cases the framework produces a stable, finite, and visually useful fit while the residual structure still shows systematic mismatch. This can occur due to external factors such as the model being an imperfect representation of system dynamics, measurement noise and more.

\section{Summary and Practical Notes}

  This chapter described a local-agent workflow for fitting ordinary differential equation models to experimental data. The workflow decomposes parameter estimation into auditable stages. A local
  Ollama-hosted coding model assists with model construction and translation, while deterministic validation, numerical checks, and preserved run directories keep the process inspectable and reproducible.

  The examples show that the workflow can handle a range of ODE fitting settings, including stiff systems, derived observables, measured forcing functions, multi-experiment fits, and custom objectives. Model choice remains important: larger or newer local coding models are generally more reliable for complex problem setup and Python-to-JAX translation.

\section*{Competing Interests}
The author has no conflicts of interest to declare that are relevant to the content of this chapter.
\bibliographystyle{unsrtnat}
\bibliography{references}

\end{document}